\documentstyle[12pt,aaspp4]{article}

\lefthead{Colbert et al.}
\righthead{Large Scale Outflows III}

\def\rosat{ROSAT}

\def\lapprox{\lesssim}
\def\gapprox{\gtrsim}
\def\etal{{et~al. }}
\def\asec{\arcsec}
\def\amin{\arcmin}
\def\han2{{H$\alpha$+[NII]}}
\def\deg{\arcdeg}

\begin{document}

\title{Large-Scale Outflows in Edge-on Seyfert Galaxies. \\
       III. Kiloparsec-Scale Soft X-ray Emission}

\author{Edward J. M. Colbert\altaffilmark{1,2,3,4},
        Stefi A. Baum\altaffilmark{1},
        Christopher P. O'Dea\altaffilmark{1},
        Sylvain Veilleux\altaffilmark{2}
        }

\altaffiltext{1}{ Space Telescope Science Institute, 3700 San Martin Drive,
                  Baltimore, MD  21218  }
\altaffiltext{2}{ Department of Astronomy, University of Maryland,
                  College Park, MD  20742 }
\altaffiltext{3}{ NAS/NRC Research Associate }
\altaffiltext{4}{ present address: Mail Code 662, Laboratory for High Energy 
                  Astrophysics, NASA Goddard Space Flight Center,
                  Greenbelt, MD  20771}

\begin{abstract}
We present ROSAT PSPC and HRI images of eight galaxies
selected from a distance-limited sample of 22 edge-on Seyfert galaxies.
Kiloparsec-scale soft X-ray nebulae extend along the galaxy minor axes
in three galaxies (NGC~2992, NGC~4388 and NGC~5506).  The extended
X-ray emission has 0.2$-$2.4 keV X-ray luminosities of 0.4$-$3.5
$\times$ 10$^{40}$ erg~s$^{-1}$.  The X-ray nebulae are 
roughly co-spatial with the large-scale radio emission,
suggesting that both are produced by large-scale galactic outflows.
Assuming pressure balance between the radio and X-ray plasmas, the 
X-ray filling factor is $\gapprox$10$^4$ times larger than the radio
plasma filling factor, suggesting that large-scale outflows in Seyfert
galaxies are predominantly winds of thermal X-ray emitting gas.
We favor an interpretation in which large-scale outflows originate as
AGN-driven jets that entrain and heat gas on kpc scales as they make
their way out of the galaxy.  AGN- and starburst-driven winds are also
possible explanations in cases where the winds are oriented
along the rotation axis of the galaxy disk.
\end{abstract}


\section{Introduction}


Observations of sub-kpc radio structures in Seyfert galaxies have shown
the presence of linear radio sources resembling jets from the active
galactic nucleus (AGN; e.g., Ulvestad \& Wilson 1989). Thus, it is
known that {\it nuclear} outflows from the AGN are present in some Seyfert
galaxies.  It is possible that, in some galaxies, these nuclear outflows
extend out of the galaxy disk and manifest themselves as {\it galactic}
outflows, perhaps as lower-power versions of the $\sim$10$-$100 kpc
scale jets in radio galaxies and quasars.  However, AGNs 
in Seyfert galaxies are weak optical and radio emitters (e.g., Antonucci 1993).
In quasars, for example, emission from the AGN dominates over
that produced by the host galaxy, but this may not be the case in some
Seyfert galaxies.  The host galaxies of Seyfert nuclei
are capable of driving powerful starburst-driven galactic outflows, similar to
those found in classical starburst galaxies (e.g., Heckman, Armus \& 
Miley 1990) if the nuclear star formation rate is high enough 
(e.g., Heckman \etal\ 1995, Heckman \etal\ 1997).
Therefore, kiloparsec scale radio structures in Seyfert galaxies
(Baum \etal\ 1993; Colbert \etal\ 1996) could be produced by galactic
outflows from a nuclear starburst or 
by nuclear outflows from the AGN.

We are currently conducting a research program to study the properties of 
these large-scale (R $\gapprox$1 kpc) outflows (LSOs) in Seyfert galaxies in 
order to investigate their nature and origin.
We have selected a distance-limited sample of 22
edge-on Seyfert galaxies (see Colbert \etal\ 1996b, hereafter paper~I)
and have searched for emission from LSOs from these galaxies 
at optical (paper~I) and radio
wavelengths (Colbert \etal\ 1996a, hereafter paper~II).  In paper~I, we
showed that optical emission suggestive of LSOs was present in
$\gapprox {{1}\over{4}}$ of all galaxies observed.  In paper~II, we
found that kiloparsec-scale radio structures were present in 
$\gapprox {{1}\over{2}}$ of all galaxies observed.  
These results show  
that LSOs are quite common in Seyfert galaxies.  
In paper~II, we also found that the 
radio structures have diffuse morphologies similar to bubbles or lobes
and that the morphologies are generally different from a spherical halo 
structure that is observed in the starburst 
galaxy M82 (Seaquist \& Odegard 1991).
In the present paper, we present ROSAT soft
X-ray (0.2 $-$ 2.4 keV) images of eight of the 22 galaxies from the 
distance-limited sample in order to study the X-ray properties of LSOs
in Seyfert galaxies.

The observational data and data reduction are
described in section 2 and the images and results are presented in section 3.
In section 4, we discuss the properties of the extended X-ray emitting gas.
Implications on the nature of LSOs in Seyfert galaxies are discussed
in section 5.  A summary is given in section 6.

\section{Observations and Data Reduction}

For ease of reference, in Table \ref{tableone}, we 
list the 22 Seyfert galaxies from
the edge-on sample (paper~I) and their assumed distances. 
The NASA HEASARC database of all pointed ROSAT
observations was used to select public archival data, taken with
either the Position-Sensitive Proportional Counter (PSPC) or
High-Resolution Imager (HRI).  
Data were available for ten of the 22 galaxies listed in Table 1.
The selected data were retrieved from the U.~S. \rosat\
archives at NASA's Goddard Space Flight Center.

A list of the archival \rosat\ data is given in Table \ref{tab6.1}.
For each observation, we list the offset of the optical position 
of the galaxy (see paper~I, Table 1) from the pointing center,
date of the observation, 
instrument used, exposure time, net counts and net count rates 
from the galaxy.  
Net counts were calculated by extracting source counts from circular regions 
of radius 1\amin\ and background counts from source-free regions surrounding 
the X-ray source (typically circular annuli).  NGC~4945 has a very large
angular size compared to the other galaxies.  For this galaxy, we used a 
rectangular source region of width 7.4\amin\ and length 21.6\amin\ oriented
along PA 42\deg.  For comparison, 
we also list (Table \ref{tab6.1}, column 8) the expected count rate for 
a hypothetical source with a
0.2$-$2.4 keV luminosity of 10$^{40}$ erg~s$^{-1}$ 
and a thermal spectrum
with temperature kT $=$ 1.0 keV with absorption from cold gas in our 
Galaxy.  This luminosity and spectrum are typical of a
starburst-driven galactic outflow (e.g., Armus \etal 1996) and of
what might be expected from an LSO.  Count rates were calculated with
the XSPEC X-ray spectral fitting software package, using the 
Galactic absorption columns listed in column 9.
Note that the hypothetical ``10$^{40}$ erg~s$^{-1}$'' LSO count rates are 
comparable in magnitude to the uncertainties in the total count rates
(column 7), so it is necessary to spatially resolve sources with 
luminosities $\sim$10$^{40}$ erg~s$^{-1}$ in order to detect them.

\placetable{tab6.1}

The point-spread functions of the PSPC and HRI are dependent on the X-ray
spectrum, but typical Gaussian core widths are 20$-$30\asec\ and 7$-$10\asec\ 
full-width half-maximum (FHWM), respectively.  Although the PSPC has poorer
resolution, it is much more sensitive than the HRI,  due to its very low 
internal background count rate.  
The energy range of both instruments is $\sim$0.2$-$2.4 keV.
Approximate total background rates for
the two instruments are 1.4  and 3.8 $\times$ 10$^{-3}$ 
counts~s$^{-1}$~arcmin$^{-2}$, for the PSPC and HRI, respectively.

It is instructive to compare the background count rate in a detect cell to
the hypothetical ``10$^{40}$ erg~s$^{-1}$'' LSO count rates listed in Table
\ref{tab6.1}, column 8.  For the PSPC, $\sim$90\% of the source counts
fall within radius 23\asec, so that we expect a background count rate of
0.06 $\times$ 10$^{-2}$ s$^{-1}$ in a detect cell.  This is much lower than
the hypothetical  ``10$^{40}$ erg~s$^{-1}$'' count rates, so the PSPC 
data are easily sensitive to LSO X-ray emission,
as long as the emission can be resolved from the nuclear source.
For the HRI, $\sim$90\% of the source counts fall within a radius of 
$\sim$10\asec, so the expected background count rate in a $\sim$10\asec\
radius detect cell is 0.03 $\times$ 10$^{-2}$ s$^{-1}$.  Again, in all cases,
10$^{40}$ erg~s$^{-1}$ emission from an LSO would be detected as long
as it is resolved from the nuclear source.

Two (NGC~4602 and NGC~7410) of the ten galaxies were not detected (count
rate $\lapprox$2$\sigma$ above the background using a 1\amin\ radius
detect cell).  By comparing the upper limits to the count rates (Table 2,
column 7) with those of the hypothetical ``10$^{40}$ erg~s$^{-1}$'' source
(Table 2, column 8), we note that the total 0.2$-$2.4 keV luminosities of 
these two galaxies are less than several $\times$ 10$^{40}$ erg~s$^{-1}$.

Images of the remaining eight galaxies were made using the PROS X-ray 
reduction software in IRAF.  First, the
raw images were binned by a factor of 8 to make new images with 4\asec\ pixels.
These images were were then smoothed with a Gaussian kernel of 
$\sigma =$24\asec.  For the HRI data, higher resolution images were
also made by smoothing with a Gaussian kernel of $\sigma =$8\asec.

In Table \ref{tab6.2}, 
we list the eight galaxies for which contour maps are presented,
along with the angular size of 1.0 kpc at the assumed distance and the figure 
number of the contour map of the image.  Note that for 
most of the galaxies, the
PSPC resolution of $\sim$25\asec\ is too poor to resolve any emission 
within $\sim$5 kpc.  The HRI resolution is significantly better.
Contour levels for all plots are in units of counts~pixel$^{-1}$ 
(225 counts~arcmin$^{-2}$).

\placetable{tab6.2}

We do not attempt to model the spectra of the X-ray data.  The relatively
low number of total counts and inadequate energy range ($\lapprox$2.4 keV)
of the PSPC detector make spectral fitting a two-component model (e.g.,
a power-law for the AGN and a thermal spectrum for the LSO) infeasible.
Although the HRI is capable of resolving extended emission from the LSO,
the pulse height amplitude (PHA) channels of the HRI are not calibrated
very well in energy, so it is not possible to estimate the temperature
of any resolved thermal emission from the HRI observations.

\section{Results}

Here we present and discuss the ROSAT HRI and PSPC soft X-ray images for
individual Seyfert galaxies in the edge-on sample.

\subsection{Individual Objects}



{\it NGC~2992.}
The HRI image of NGC~2992 was previously published by Weaver \etal\ (1996),
but they did not note the presence of any extended X-ray emission.
In Figure 1a, 
we show a contour map of the HRI image of NGC~2992.
It is immediately obvious that 
the companion galaxy NGC~2993 is also 
a source of soft X-ray emission.  It is
also clear that soft X-ray emission extends out to $\sim$1\amin\ ($\sim$9 kpc)
to the southeast in PA $\sim$100\deg, roughly along the galaxy minor axis.
%
Although the PSPC image (not shown) 
is much more sensitive to faint emission, the strong
emission from the nuclear source dominates the image and thus
no extended emission is resolved.

\placefigure{xfig1new}


{\it NGC~4235.}
The X-ray source in both (1992 June and 1992 December) PSPC images of
NGC~4235 is unresolved.
The PSPC 
resolution of $\sim$25\asec\ corresponds to $\sim$3.9 kpc and the X-ray 
emission from the nuclear source certainly dominates over any possible 
extended X-ray emission on kpc scales.



{\it NGC~4388.}
The HRI image of NGC~4388 was previously published by Matt \etal\ (1994).
These authors found extended emission out to radii $\sim$45\asec\ 
(7.3 kpc), roughly evenly distrubuted over all azimuthal angles.
Contour maps of the HRI and PSPC images are shown in Figures 1b and 1c,
respectively.  
In the PSPC image, we note the presence of a larger scale X-ray
structure, extending out to radii $\sim$2\amin\ (19.5 kpc) in P.A. 
$\sim$65\deg.



{\it NGC~4602.}
A field near NGC~4602 was observed with the PSPC, but 
NGC~4602 was not detected.  Unfortunately, during the observation, NGC~4602 
was offsect $\sim$19.2\amin\ from the center of the field of view, which is
near the support ring on the PSPC detector.  


{\it NGC~4945.}
In Figure 1d, 
we show a contour plot of the 1992 August PSPC image of NGC~4945,
overlaid on a grayscale plot of the optical emission.
NGC~4945 is very near the Galactic
plane ($b \approx$ 13\deg) and, consequently, the Galactic absorbing column
is quite large (N$_H \sim$ 1.5 $\times$ 10$^{21}$ cm$^{-2}$, Table 2).
Thus, soft X-ray emission
below $\sim$1$-$2 keV will be severely absorbed in the image.
No extended soft X-ray emission is evident in the 1993 July PSPC image either.



{\it IC~4329A.}
This galaxy is located at a distance of $\sim$64 Mpc, at which the PSPC 
resolution of $\sim$25\asec\ corresponds to $\sim$7.8 kpc. More complications
arise because this type~1 Seyfert nucleus is a very luminous X-ray source and 
photons within radii $\sim$ 2\amin\ (corresponding to physical distances
$\lapprox$35 kpc) are dominated by the unresolved nuclear source.
A contour plot of the PSPC image is shown in Figure 2a. 

\placefigure{xfig5new}


{\it NGC~5506.}
In Figure 2b, 
we show a contour map of a PSPC image of NGC~5506.  Note
that it appears to be extended along P.A. $\sim$150\deg, roughly the same
orientation as the large-scale radio emission (paper~II).
On smaller scales, as shown in the HRI image (Figure 2c), the emission is 
extended in roughly the same P.A.



{\it ESO~103-G35.}
The galaxy ESO~103-G35 was observed with 
the PSPC, but very few ($\sim$73) net counts 
were detected. 
We do not find any evidence for extended X-ray emission.



{\it NGC~7410.}
The PSPC observation of the galaxy NGC~7410 was quite short 
($\sim$1 ksec) and there
were only 7.4$\pm$5.5 net counts from the galaxy, too weak to claim a definite
detection.


{\it NGC~7590.}
The PSPC observation of NGC~7590 was also quite short and we do not
find any evidence for extended emission in the image. 


\subsection{Suitability of Imaging Data for Detecting Emission from
LSOs}

As mentioned in section 2, all of the images are sensitive to soft 
X-ray luminosities of $\sim$10$^{40}$ erg~s$^{-1}$, comparable to what
would be expected from an LSO.
Thus, the only issue
with respect to the suitability of the data for detecting extended emission
from LSOs with luminosities $\gapprox$10$^{40}$ erg~s$^{-1}$ 
would be whether the
emission can be resolved from the bright nuclear X-ray source.  As can
be seen from the angular scales listed in Table \ref{tab6.2}, the
PSPC image of only one galaxy (NGC~4945) 
has spatial resolution suitable for resolving sources
on scales $\sim$1$-$2 kpc.  The three HRI images of
NGC~2992, NGC~4388 and NGC~5506 also have suitable resolution for
imaging emission on scales $\gapprox$1 kpc.

Although the spatial resolution of the PSPC images is $\sim$3$-$5 times
worse than the HRI, the PSPC images are useful for searching for very
extended X-ray emission, such as that in NGC~4388.  For example,
the PSPC images of NGC~2992, NGC~4235, NGC~4388, NGC~4945, NGC~5506 and 
NGC~7590 are sensitive to emission on size scales $\gapprox$3$-$5 kpc.

In summary, ROSAT imaging data suitable for searching for emission from
LSOs on scales $\gapprox$1 kpc were available for four of the eight galaxies,
and data suitable for searching for emission $\gapprox$3$-$5 kpc were 
available for six of the eight galaxies.  For comparison, large-scale
radio structures from LSOs have typical radii of $\sim$1$-$5 kpc (paper~II).

\subsection{Extended Emission in NGC~2992, NGC~4388 and NGC~5506}

As mentioned in section 3.1, three (NGC~2992, NGC~4388 and NGC~5506)
of the galaxies show extended emission at radii $\gapprox$ 1 kpc.
Ideally,
one would like to make an image of the extended
emission by subtracting the unresolved nuclear source.  Unfortunately, this
is not a straightforward procedure as the point-spread functions for
both instruments are dependent on the energy spectrum of the source 
(especially for the PSPC).  In addition, especially
important for the HRI, an uncertainty in the aspect solution returned by 
the spacecraft introduces non-axisymmetric errors on size scales 
of $\sim$5$-$10\asec.  Subtraction of 
the nuclear source could be attempted if there is a bright 
unresolved X-ray source 
near the pointing center that can be used as a 
point-spread function, but such
sources were not present in the ROSAT images of these galaxies.  Aspect
errors can be reduced using sophisticated image reconstruction methods
(e.g., Morse 1994); however, these methods were not found to be
very effective for these 
observations, due to the relatively low number of total counts.  
In the following subsections, we describe measurements of the net counts of 
the extended soft X-ray emission in these three galaxies.

\subsubsection{NGC~2992}

The extended emission in NGC~2992 (Figure 1a) 
extends along PA 112\deg, but
does not appear extended along other PAs.  
The surface brightness of the extended emission in the HRI image is only
slightly brighter than that from the wings of the unresolved nuclear
source.  In order to make evident the faint extended emission, in
Figure \ref{xfig9},
we show radial profiles of the X-ray emission in four quadrants
(PA $=$ -30 $-$ 60\deg, 60 $-$ 150\deg, 150 $-$ 240\deg, and 240 $-$330\deg).
The radial profile for the second quadrant (as listed above) is significantly
peaked above the others (which serve as model PSFs)
between $\sim$35\asec\ and 45\asec.  This excess is
noticeable in radial profile plots binned at either 10 or 20 photons per
radial bin, i.e., it is not a chance surplus of photons caused by binning
the data (see Figure \ref{xfig9}).
For example, between radii 35\asec\ and 45\asec,
there are 20 total counts in the second quadrant and 27 total
counts from {\it all three other quadrants together}.  After 
background subtraction,
the net number of counts of this extended emission in the second quadrant is 
11.0$\pm$4.8.

\placefigure{xfig9}

\subsubsection{NGC~4388}

As noted in Table \ref{tab6.1}, relatively few source counts were detected 
in the data.  The HRI image (Figure 1b) 
appears extended, 
but only $\sim$150 counts
are present in the raw image.  The PSPC image also shows extended emission,
is more sensitive, and contains more counts in the raw image. We use
this image to estimate the count rate of the extended emission.  
In order to verify that the extended emission was not due to `ghost
images' of very soft (E $\lapprox$ 0.2 keV) photons (Nousek \& Lesser
1993), we omitted photons from PI channels 16 (of 256) and below.
The total 
number of counts within a circle of radius 1\amin\ (which encloses $>$95\% of
the total counts from a point source) is 441. 
The total counts 
within a tilted rectangular region of width 2.5\amin\ and 
length 4.1\amin, oriented along PA 46.5\deg (enclosing the whole galaxy)
is 784, 
a difference of 343 
counts.  After subtraction of the background
photons, this becomes 113 
net counts.
Although the central source may be contributing some of the photons
beyond radii 1\amin, this amounts to
less than 5\% of 339 
net (background subtracted) 
counts, or less than 17 photons.  Taking this 
uncertainty into account, the net counts and uncertainty of the extended 
X-ray emission is 113$\pm$28.\footnotemark[1]            
\footnotetext[1]{
  Approximately the same number of net photons was found if all PI
  channels were used (114$\pm$29), indicating that `ghost images' from
  very soft photons are not a significant factor in producing the
  observed  extended emission.
}

\subsubsection{NGC~5506}

As can be seen in Figure 2c, 
the extended emission surrounding the central
X-ray source in NGC~5506 is oriented along PA $\sim$160\deg, but has a rather
large opening angle.  In Figure \ref{xfig11},
we show a radial profile of the counts
per unit area, binned so there are at least 20 photons for each data point.
Since the extended emission in NGC~5506 is directed over a large range
in PA, we cannot easily use radial profiles in other PA ranges to model
the PSF (as we did for NGC~2992).  Instead, for
comparison, we also plot the radial profile from a deep observation 
(106,359 net counts) of the point source HZ43, taken from a ROSAT HRI
calibration observation (ROSAT sequence number RH141873N00).  The 
HZ43 profile has been normalized to the
NGC~5506 profile by dividing by the ratio of the total number 
of counts~area$^{-1}$~time$^{-1}$ for NGC~5506 to that of HZ43.
The area used was that within radius 10\arcsec.

\placefigure{xfig11}

NGC~5506 begins to show an increased X-ray surface brightness above the 
HZ43 PSF beyond radius $\sim$5\asec.  As noted above, the aspect solutions 
for HRI observations can produce errors on $\sim$10\asec\ scales, so it
it is not clear whether the apparent extent between 5 and 10\asec\ is real.
However, the excess emission becomes even
more pronounced between radii $\sim$10\asec\ and 25\asec, which is not
due to improper aspect solution.  A conservative estimate of the net 
extended counts, calculated by subtracting the total counts from NGC~5506
between 10 and 25\asec\ from that of the normalized HZ43 distribution,
is 72$\pm$13. 

\subsubsection{X-ray Luminosities of the Extended Soft X-ray Emission}

In Table \ref{tab6.3},
we list ranges of the size of the extended emission in the
three galaxies.  Also listed is the range in PA over which the emission is 
extended, the
net counts, and the net count rates of the extended emission.

\placetable{tab6.3}

Estimating fluxes and luminosities of the extended emission requires 
knowledge about the intrinsic spectrum of the emission.
We assume a thermal spectrum (Raymond \& Smith 1977) with absorption due
to cold gas in our Galaxy.  In Table \ref{tab7.3}, column 4, we list 0.2$-$2.4
keV luminosities of the extended emission for assumed thermal spectra 
for two sample plasma temperatures: (1) kT $=$  0.1 and (2) kT $=$ 1.0 keV.
We have used the X-ray spectral fitting program XSPEC to calculate X-ray
fluxes from PSPC and HRI count rates, using the Galactic absorption
columns listed in Table 2.

%

\placetable{tab7.3}


%


\section{Nature of Extended Soft X-ray Emission}

\subsection{Emission Mechanism}

We explore several possibilities for the emission mechanism producing 
the observed soft X-rays.

\subsubsection{Scattered Nuclear Emission}

The extended soft X-ray emission 
originates from regions $\gapprox$1 kpc from the nucleus, so it
is not likely to be scattered X-rays from the nuclear X-ray source.  
The electron-scattering region in Seyfert galaxies is estimated to be
$\sim$ 1 pc (e.g., Krolik \& Begelman 1986).

\subsubsection{Synchrotron Emission}

In bright jets such as that in M87, soft (E $\sim$ 1 keV) 
X-ray emission from knots or hotspots is 
consistent with synchrotron emission (e.g., Biretta 1993).  The expected 
synchrotron luminosity in the ROSAT (0.2$-$2.4 keV) band from a synchrotron
source with 6~cm radio luminosities comparable to the large-scale radio
structures in Seyfert galaxies
($\sim$10$^{21}$ W~Hz$^{-1}$; paper~II) is
$\sim$10$^{40}$ erg~s$^{-1}$
(assuming the synchrotron luminosity L$_\nu \propto \nu^{-0.7}$),
which is comparable to the observed ROSAT X-ray luminosities.
However,
the lifetime of the relativistic electrons which would produce such
high energy X-ray synchrotron emission is much shorter than that of 
electrons producing cm-wave radio emission.  Specifically, the lifetimes
are a factor [$\nu$(X-ray)/$\nu$(radio)]$^{0.5}$ shorter.  This means
that the electrons emitting 1~keV synchrotron emission live $\sim$10$^{-4}$
as long as the ones emitting at 6~cm, or $\sim$10$^3$ yrs,
based on the 6~cm radio synchrotron lifetimes of the large-scale radio
structures (paper~II).  Since there is no evidence on kiloparsec scales 
for a source that would supply relativistic electrons (e.g., a jet), we
disregard synchrotron emission as a possible emission mechanism.

\subsubsection{X-ray Point Sources in the Galaxy Halos}

Since the X-ray luminosities of the extended emission is $>$~10$^{40}$
erg~s$^{-1}$ (Table 5), over 100 X-ray binaries emitting at 10$^{38}$
erg~s$^{-1}$ would be required to 
provide enough X-ray emission.  Such a large number of very luminous X-ray 
binaries is not expected in a galaxy halo.  Although some X-ray
supernovae emit with X-ray luminosities $\sim$10$^{40}$ erg~s$^{-1}$
(e.g., Ryder \etal\ 1993), such objects are quite rare.  The fact
that the extended X-ray emission is roughly co-spatial with large-scale
diffuse radio emission (section 4.3) argues that hot plasma in an LSO is
producing extended, {\it diffuse} X-ray emission.

\subsubsection{Thermal Emission from a Hot Plasma}

Gas at temperatures $\sim$ 10$^6$$-$10$^7$ K emits soft X-rays in 
an energy range in which the ROSAT detectors are quite sensitive.
Our present ROSAT data are not suitable for measuring the temperature of the 
extended X-ray emission, so we have not attempted
to do so.  A thermal plasma with a 
temperature kT as low as $\sim$0.1 keV or as high as several keV could be 
producing the soft X-rays.  Wind shocks could heat gas to these temperatures.  
The temperature of gas which has been heated by a shock with velocity $v$ is
$$kT = {{3}\over{16}} \mu m_H v^2,$$
where $\mu$ is the mean atomic weight of the gas 
and $m_H$ is the mass of
the hydrogen atom.  For shocks with  velocities $\sim$100$-$200 km~s$^{-1}$
the shocked gas will emit at temperatures $\lapprox$0.05 keV, which is too
low to be detected with ROSAT. However, wind shocks with velocities $\gapprox$
300$-$1000 km~s$^{-1}$ will heat gas to temperatures of $\sim$0.1$-$1.0 keV, 
which is in the ROSAT bandpass.
An example of extended soft X-ray emission from a thermal plasma occurs
in the Seyfert galaxy NGC~2110, in which the plasma temperature is 
$\approx$ 1 keV (Weaver \etal\ 1995).
Since the extended emission in NGC~2992, NGC~4388 and NGC~5506 is
positioned out of the galaxy disk (and therefore is not likely to be
emission from sources in the galaxy disk), we favor an interpretation as 
thermal emission  from a hot plasma with temperature
$\sim$10$^6$$-$10$^7$ K (kT $\sim$ 0.1 $-$ 1.0 keV).

\subsection{Derived Properties}

In Table \ref{tab7.3}, for each of the three galaxies with extended
emission,
we list important properties of the extended X-ray gas for 
two sample plasma temperatures: kT $=$ 0.1 and 1.0 keV.  The soft X-ray
luminosity L$_X$ in the ROSAT band (0.2$-$2.4 keV)
is given by 
$$ L_X = \int n^2 \Lambda_X dV,$$
where n and V are the density and volume of the emitting gas and
$\Lambda_X$ is the emissivity of a thermal plasma (Raymond \& Smith 1977) for 
the ROSAT bandpass.  The X-ray emissivity is nearly constant over the 
ROSAT energy bandpass and we have used a value of 
$\Lambda_X =$ 10$^{-22.4}$ erg~cm$^3$~s$^{-1}$, as given in 
Suchkov \etal\ (1994).  The volume (Table \ref{tab7.3}, column 2) of the 
emitting region
was calculated assuming a conical region with inner and outer radii and opening
angle as given in Table 4.
$$ V = {{2\pi}\over{3}} [1-\cos\alpha] 
        (R_{outer}^3 - R_{inner}^3),$$
where $\alpha$ is the half-angle of the cone.

We assume the hot plasma fills the emitting volume with filling factor $f$
and list values of $fn^2V = L_X/\Lambda_X$ in Table \ref{tab7.3}, column 5.
We then calculated densities 
(f$^{{{1}\over{2}}}$n $=$ (L$_X$/$\Lambda_X$ V)$^{{{1}\over{2}}}$; column 6),
pressures (f$^{{{1}\over{2}}}$2nkT; column 7),
and cooling times of the hot gas.  The cooling times have been
calculated from the following formula:
$$ t_{cool} \approx {{nkT fV}\over{L_X}} = 
   kT f^{{{1}\over{2}}} ({{V}\over{\Lambda_X L_X}})^{{{1}\over{2}}}.$$



\subsection{Associated Radio Emission}



In all three galaxies with extended X-ray emission,
there is strong evidence for a large-scale outflow
from both radio and optical observations (papers I and II).
It is also noteworthy
that the extended X-ray nebulae are roughly cospatial with the extended
radio emission (see Figures \ref{ch7fig3} and \ref{ch7fig4}),
which is further evidence that the X-ray 
emission is produced by the galactic outflow.  

\placefigure{ch7fig3}

\placefigure{ch7fig4}

\section{Nature of Large-scale Outflows in Seyfert Galaxies}

\subsection{Frequency of Occurrence}

Of the four galaxies in our sample for which ROSAT images with suitable 
sensitivity and $\sim$1 kpc spatial resolution were available (see section
3.2), kpc-scale soft X-ray nebulae were found
in three (NGC~2992, NGC~4388 and NGC~5506) of the galaxies.  
Although the eight objects observed make up an unbiased subsample of the
distance-limited sample, the sample of four objects for which suitable
deep observations were made are probably biased.  All
three of these galaxies were previously known to have 
extended optical and radio emission.  
Thus, the 75\% detection rate for extended X-ray emission may be an 
overestimate.
%
However, it
is noteworthy that 
the deep ROSAT observations strongly reinforce the
presence of LSOs in these galaxies, confirming the evidence provided by 
optical and radio observations.

Wilson (1994) shows that at least four of six Seyfert galaxies studied
in their research program show spatially extended soft X-ray emission.
This suggest that 
extended soft X-ray emission from
LSOs is common in Seyfert galaxies.  Since LSOs are so common, the soft X-ray 
emission from LSOs could, in some cases,  explain the ``soft excess'' 
component observed in total aperture X-ray spectra of Seyfert galaxies,
which is present in $\gapprox$50\% of Seyfert galaxies 
(e.g., Turner \etal\ 1991).

A systematic study of the frequency of occurrence of X-ray halos in
edge-on normal and starburst galaxies has not been done.  
ROSAT X-ray luminosities of halos that have been detected in normal and
starburst galaxies are $\sim$ several $\times$ 10$^{39}$ erg~s$^{-1}$
(see, e.g., Bregman \& Houck 1997), which is a factor of
$\sim$10 or more smaller than the luminosities of the extended emission
we find in our study of Seyferts.  This may indicate that there are two
different types of outflows occurring.
It would be interesting to compare the results of a study of normal and
starbursting galaxies with our results for Seyfert galaxies.



\subsection{Physical State of the Outflowing Gas}


We can investigate the state of the gas in the LSOs by comparing the
implied pressures and filling factors of the optically-emitting gas, the
radio plasma, and the X-ray plasma.  First we compare the optical and 
X-ray emitting gas.
%
Optical line ratios from emission-line clouds at distances $\lapprox$1 kpc 
from the Seyfert nucleus 
(in the so-called extended narrow-line region [ENLR]) 
suggest that the clouds
have electron densities
n$_e \sim$ 10$^2$$-$10$^3$ cm$^{-3}$ 
and temperatures T$_e$ $\approx$ 10$^4$ K (e.g., Robinson 1989, 
Robinson 1994), implying pressures n$_e$kT $\sim$ 
10$^{-10}$$-$10$^{-9}$ dyne~cm$^{-2}$.  
We found that the extended X-ray emission from
NGC~2992, NGC~4388 and NGC~5506 is located at larger radii 
(between $\sim$2 and 13 kpc.)
Since the pressure of outflowing wind gas is expected to drop as $R^{-2}$,
the pressures of hot X-ray gas at the smaller radii of the 
ENLR clouds ($\sim$1 kpc) will be higher than those listed in Table 5.
For example, LSO gas with pressures 
$\sim$ 10$^{-12}$$-$10$^{-10.1}$  $f^{-{{1}\over{2}}}$ dyne~cm$^{-2}$ 
(the range for the three galaxies listed above; Table 5) at a sample
radius of 5 kpc will have higher pressures $\sim$10$^{-10.6}$$-$10$^{-8.7}$
$f^{-{{1}\over{2}}}$ 
dyne~cm$^{-2}$ at $\sim$1 kpc.  These pressures are roughly consistent
with the electron pressures of the ENLR gas at the same radii
($\sim$ 10$^{-10}$$-$10$^{-9}$ dyne~cm$^{-2}$).
Thus, if the X-ray gas and the ENLR clouds are in pressure equilibrium,
then the X-ray filling factor $f \sim$ 0.1$-$1.

Next, we compare the radio-emitting plasma with the X-ray plasma.
The minimum-energy magnetic
pressures of the large-scale radio structures 
imply that the kpc-scale {\it radio plasma} 
has pressures 
$\sim$10$^{-14}$$-$10$^{-13}$ $\phi^{-{{4}\over{7}}}$ dyne~cm$^{-2}$,
where $\phi$ is the volume filling factor of the radio plasma (paper~II).
Assuming pressure balance between the X-ray and radio-emitting plasmas
(a reasonable assumption since the large-scale X-ray and radio structures
are roughly co-spatial), the ratio of filling factors
$\phi/f \lapprox$ 10$^{-4}$,
i.e., there is heavy mixing of thermal gas within the radio-emitting
volume or the radio plasma is confined to filamentary (or shell-like)
regions.  


In summary, 
our results suggest that the outflowing material in the LSOs of Seyfert
galaxies
is primarily composed of hot, thermal X-ray emitting gas (with only
a small contribution of synchrotron-emitting relativistic electrons).
The radio and X-ray plasmas are assumed to be 
in rough pressure equilibrium with the optical ENLR clouds.
Krolik \& Vrtilek (1984) describe a scenario in which
the ENLR clouds in fact condense out of outflowing wind gas and
create a two-phase medium in the ENLR, which is consistent with our results.

%

\subsection{Possible Scenarios for LSOs in Seyfert Galaxies}

In paper~II, we noted that the large-scale radio structures in galaxies
in our sample had diffuse morphologies, but that the morphologies did
not resemble radio halos in starburst galaxies (like M82).  
We concluded that LSOs are most easily understood as AGN-driven outflows
that are somehow diverted away from the galaxy disk on scales
$\gapprox$1 kpc.  Various
mechanisms (e.g., cloud collisions, bending by ram pressure forces, and
buoyancy forces) may act to divert the outflows as they blow 
outward from the nuclear region.  


In the present paper, we find that LSOs are primarily composed of hot,
X-ray emitting plasma, which suggests that, on kpc-scales, the outflows
are more similar to wide-angled winds than to radio plasma jets from
the AGN.  A possible explanation is that AGN-driven jets 
entrain large amounts of gas as they make their way
out of the galaxy.  The entrained gas could be shock-heated to 
temperatures $\gapprox$10$^6$ K by winds with velocities 
$\gapprox$300 km~s$^{-1}$.

Another possibility is that the outflows originate as winds very close
to the nucleus.  If the jets are stopped at small radii from the
nucleus,
the kinetic energy of the jet may become thermalized and heat the 
gas in the nuclear region (as does a compact starburst for a 
starburst-driven wind, see Heckman, Armus \& Miley 1990).
The high-pressure region can then blow out of the galaxy disk in a wind.
A nuclear starburst can also power a galactic wind in a Seyfert galaxy,
if the nuclear star-formation rate is high enough.
In the wind scenario, the large-scale X-ray and radio structures
would tend to be oriented
perpendicular to the galaxy major axis, which is the case for some LSO
Seyfert galaxies (e.g., NGC~4388).


Detailed studies are needed to 
determine which energy source (AGN or nuclear starburst) dominates the 
energy input to the LSO in any given galaxy.

\section{Summary}

ROSAT PSPC and HRI soft X-ray imaging data for ten
of the 22 galaxies from the complete sample were retrieved from the ROSAT
archives and analyzed.
Eight of these galaxies were detected.  Suitable data for detecting 
extended soft (0.2$-$2.4 keV) X-ray emission on size scales $\gapprox$1
kpc were available for
four galaxies and extended soft X-ray emission was found in three 
(NGC~2992, NGC~4388 and NGC~5506) of these four galaxies.
The extended soft X-ray emission 
has 0.2$-$2.4 keV X-ray luminosities
between 0.4 and 3.5 $\times$ 10$^{40}$ erg~$^{-1}$.  Each one of these three 
galaxies also has
strong evidence for a large-scale outflow from previous optical and/or
radio observations (papers I and II).
The large-scale radio, optical and X-ray emission is roughly
co-spatial, and
the soft X-ray emission is most plausibly explained by thermal emission
from a gas in a galactic outflow.
Assuming pressure balance between the radio and X-ray plasmas, the
filling factor of the X-ray emitting gas is $\gapprox$10$^4$ times
larger than
that of the radio plasma.
Large-scale outflows in Seyfert galaxies are most easily explained as 
wide-angled winds (as opposed to collimated jets) of hot, X-ray emitting
gas that blow away from the 
nuclear region.
We favor an interpretation in which LSOs originate as AGN-driven jets
that entrain and heat gas on kpc scales as they make their way out of
the galaxy.  An alternate explanation is that LSOs originate as winds
on small scales and this could be occurring in galaxies in which the
LSOs are oriented roughly perpendicular to the galaxy major axis.

\acknowledgments

E.J.M.C. thanks Kim Weaver, Michael Dahlem and Martin Elvis for providing
information and for helpful discussions and Michael Corcoran for helping
with the NASA/HEASARC data archive.  E.J.M.C. acknowledges support from
the Director's Office of the Space Telescope Science Institute, NASA
grant NAG5-3016, and the National Research Council.
S.V. acknowledges the financial support of NASA through grant LTSA
035-97.
This research has made extensive use of the NASA/IPAC
Extragalactic Database (NED), which is operated by the Jet Propulsion
Laboratory, Caltech, under contract with NASA.  The Digitized Sky
Surveys were produced at STScI under U.S. Government grant NAGW-2166.
This paper represents a
portion of E.J.M.C.'s Ph.D. thesis, submitted in partial fulfillment
of the requirements of the Graduate School of the University of Maryland.

\clearpage

\begin{deluxetable}{lclr}
\scriptsize
\tablewidth{5in}
\tablecaption{ {\bf Complete Statistical Sample of Edge-on Seyfert Galaxies}
  \label{tableone} }
\tablehead{
\colhead{Galaxy} & 
\colhead{ROSAT} &
\colhead{Seyfert} &
\colhead{D\tablenotemark{a}} \\
\colhead{Name} & 
\colhead{Data?} &
\colhead{Type} &
\colhead{(Mpc)} \\
} 
\startdata
IC~1657       &     & 2   & 47.4 \\ 
UM~319        &     & 2   & 63.1 \\ 
Mrk~993       &     & 2   & 62.1 \\ 
Mrk~577       &     & 2   & 69.1 \\
Ark~79        &     & 2   & 68.8 \\ 
NGC~931       &     & 1   & 66.6 \\ 
NGC~1320      &     & 2   & 36.0 \\ 
NGC~1386      &     & 2   & 20.0 \\ 
NGC~2992      & Yes & 2   & 30.9 \\ 
MCG~$-$2-27-9 &     & 2   & 62.0 \\ 
NGC~4235      & Yes & 1   & 32.1 \\ 
NGC~4388      & Yes & 2   & 33.6 \\ 
NGC~4602      & Yes & 1.9 & 34.0 \\ 
NGC~4945      & Yes & 2   &  6.7 \\ 
IC~4329A      & Yes & 1   & 63.9 \\ 
NGC~5506      & Yes & 2   & 24.2 \\ 
ESO~103-G35   & Yes & 2   & 53.1 \\ 
NGC~6810      &     & 2   & 26.1 \\ 
IC~1368       &     & 2   & 52.2 \\ 
IC~1417       &     & 2   & 57.5 \\ 
NGC~7410      & Yes & 2   & 23.3 \\ 
NGC~7590      & Yes & 2   & 21.3 \\ 
\enddata
\tablenotetext{}{Optical positions, axial ratios and recessional velocities
for these objects are listed in paper~I, Table 1. }
\tablenotetext{a}{Assumed distance in Mpc.  Except for NGC~1386 and NGC~4945,
  distances were calculated from recessional velocities listed in RC3
  (see paper~I), using H$_o =$ 75.  The distance to NGC~1386 was taken as
  20.0 Mpc (the distance to the Fornax cluster) and a distance of 6.7 Mpc was
  assumed for NGC~4945.}
\end{deluxetable}

\clearpage

\begin{deluxetable}{lrllrrccc}
\scriptsize
\tablecaption{ {\bf ROSAT Archival Data for Edge-on Seyfert Galaxies } 
  \label{tab6.1} }
\tablewidth{6.1in}
\tablehead{
\colhead{Galaxy} & \colhead{Offset\tablenotemark{a}} & 
\colhead{Obs Date} & \colhead{Instr} & 
\colhead{Exp.\tablenotemark{b}} & 
\colhead{Cnts\tablenotemark{c}} & \colhead{Count\tablenotemark{c}} & 
\colhead{10$^{40}$ erg~s$^{-1}$} &
\colhead{N$_H^{Gal,}$\tablenotemark{d}} \\
\colhead{Name} & \colhead{(arcmin)} & 
\colhead{(YYMMDD)} & \colhead{} & 
\colhead{Time} & \colhead{} & 
\colhead{Rate} & \colhead{Cnt Rate\tablenotemark{d}} \\
\colhead{} & \colhead{} & 
\colhead{} & \colhead{} & \colhead{(s)} & 
\colhead{} & \colhead{(10$^{-2}$}&
\colhead{(10$^{-2}$}&
\colhead{(10$^{20}$} \\
\colhead{} & \colhead{} & 
\colhead{} & \colhead{} & \colhead{} &
\colhead{} & \colhead{s$^{-1}$)} & 
\colhead{s$^{-1}$)} &
\colhead{cm$^{-2}$)} \\
\colhead{(1)} & \colhead{(2)} & \colhead{(3)} & \colhead{(4)} & \colhead{(5)} &
\colhead{(6)} & \colhead{(7)} & \colhead{(8)} & \colhead{(9)} \\
} 
\startdata
NGC~2992  &  0.3 & 931115$-$29 & PSPC & 
  18602 & 1781$\pm$44  & 9.6$\pm$0.2 & 0.7 & 5.5 \\ 
&            0.3 & 920523      & HRI  & 
  10840 &  276$\pm$23  & 2.6$\pm$0.2 & 0.3 & \\ 
NGC~4235  &  0.1 & 920701$-$07 & PSPC &  
   9142 & 1265$\pm$38 & 13.8$\pm$0.4 & 0.8 & 1.6 \\ 
&            0.1 & 921208$-$23 & PSPC & 
  10297  & 1819$\pm$45 & 17.7$\pm$0.4 & 0.8 & \\ 
NGC~4388  &  0.5 & 930616$-$0707 & PSPC & 
  11650 & 344$\pm$23 & 3.0$\pm$0.2 & 0.7 & 3 \\ 
&            0.3 & 911207$-$09& HRI  & 
  11256 & 147$\pm$19 & 1.3$\pm$0.2 & 0.3 & \\ 
NGC~4602  & 19.3 & 920714  & PSPC &  
   1261 &   & $<$1.0 & 0.7 & 2.3  \\ 
NGC~4945  &  0.1 & 920812$-$15 & PSPC & 14190 & 
  1426$\pm$66 & 10.1$\pm$0.5 & 11.6 & 14.5 \\ 
&            0.1 & 930710$-$11 & PSPC &  9028 & 
  1202$\pm$53 & 13.3$\pm$0.6 & 11.6 & \\ 
IC~4329A  &  0.1 & 930114 & PSPC &  
   8230 & 22324$\pm$151 & 271$\pm$2 & 0.2 & 4.6 \\ 
NGC~5506  &  0.3 & 920124$-$27& PSPC & 
   4462 & 1209$\pm$36 & 27.1$\pm$0.8 & 1.2 & 4.2 \\ 
&           0.3 & 940710$-$28 & HRI  & 
  18019 & 866$\pm$36 & 4.8$\pm$0.2 & 0.5 & \\ 
ESO~103$-$G35 & 4.6 & 930331$-$0413 & PSPC & 
  16813 & 73$\pm$13 & 0.4$\pm$0.1 &  0.2 & 7.6\\ 
NGC~7410  &  0.2 & 931203 & PSPC &  
  1491 &  & $<$1.1 & 1.6 & 1.3  \\ 
NGC~7590  & 10.1 & 930504$-$05 & PSPC &  
  7241 & 107$\pm$14  & 1.5$\pm$0.2 & 1.8  & 2.0  \\ 
\enddata
\tablenotetext{a}{Difference between optical position of galaxy (see
  paper I) and pointing center of observation}
\tablenotetext{b}{Net exposure time of observation}
\tablenotetext{c}{Net (background-subtracted) counts and count rates from
  the galaxy (see text for more details)}
\tablenotetext{d}{ Sample count rates (for comparison) of a hypothetical 
source with a 0.2$-$2.4 keV luminosity of 10$^{40}$ erg~s$^{-1}$, located 
at the distance of the galaxy.  We have assumed a thermal spectrum and 
have taken into account absorption of soft X-rays due to the
the Galactic column toward the galaxy.  Galactic columns (column 9) were 
taken from Elvis, Lockman, \& Wilkes (1989), Boller et al. 1992, Iwasawa
et al. 1996, and Dickey \& Lockman 1990. }
\end{deluxetable}

\clearpage

\begin{deluxetable} {lrcl}
\tablecaption{{\bf X-ray Images of Edge-on Seyfert Galaxies}}
\tablehead{
\colhead{Galaxy} & \colhead{Scale\tablenotemark{a}} &
\colhead{Fig No.\tablenotemark{b}} &
\colhead{Comment\tablenotemark{b}} \\
\colhead{Name} & \colhead{(kpc~arcmin$^{-1}$)} & \colhead{} & \colhead{} \\
} 
\startdata
  NGC~2992      & 9.00  & 1a & HRI extended \\
  NGC~4235      & 9.36  & {...} & PSPC only, unresolved \\
  NGC~4388      & 9.78  & 1b,c & PSPC extended \\
  NGC~4945      & 1.95  & 1d & PSPC only, no extended emission\tablenotemark{c} \\
  IC~4329A      & 18.60 & 2a & PSPC only,  unresolved \\
  NGC~5506      & 7.02  & 2b,c & Extended \\
  ESO~103-G35   & 15.42 & {...} & PSPC only, weak nucleus, unresolved\\
  NGC~7590      & 6.18  & {...} & PSPC only, weak nucleus, unresolved \\
\enddata
\tablenotetext{a}{Linear scale using assumed distances listed in Table 
  \ref{tableone}}
\tablenotetext{b}{Figure number for contour plots of ROSAT images and
  comment about whether extended emission is apparent in the image}
\tablenotetext{c}{The Galactic column toward NGC~4945 is very high
(N$_H$ $\gapprox$ 10$^{21}$ cm$^{-2}$) and any extended soft X-ray emission 
  with 
temperatures $\lapprox$1$-$2 keV would be significantly absorbed.}
\label{tab6.2}
\end{deluxetable}

\clearpage

\begin{deluxetable} {lccrclcc}
\tablecaption{{\bf Dimensions and Count Rates of Extended X-ray Emission }}
\tablehead{
\colhead{Galaxy} & \multicolumn{2}{c}{Radius\tablenotemark{a}} & 
\multicolumn{2}{c}{Orientation\tablenotemark{b}} &
\colhead{Instr} & 
\colhead{Cnts\tablenotemark{c}} & \colhead{Cnt Rate\tablenotemark{c}} \\
\colhead{} & \colhead{} & \colhead{} & \colhead{P.A.} & \colhead{Half} &
\colhead{} & \colhead{} & \colhead{} \\
\colhead{} & \colhead{} & \colhead{} & \colhead{} & \colhead{Angle} &
\colhead{} & \colhead{} & \colhead{} \\
\colhead{} & \colhead{(arcsec)} & \colhead{(kpc)} & 
\colhead{(deg)} & \colhead{(deg)} &
\colhead{} & \colhead{} & \colhead{(10$^{-2}$ cnts s$^{-1}$)} \\
} 
\startdata
  NGC~2992 & 35$-$45 & 5.3$-$6.8 & 110 & 10 & HRI  & 11$\pm$5 & 0.10$\pm$0.04 \\
  NGC~4388 & $<$85   & $<$13.6   & 65 & 65 & PSPC & 113$\pm$28   & 1.0$\pm$0.2 \\
  NGC~5506 & 10$-$25 & 1.2$-$3.0 & $-$20 & 45 & HRI & 72$\pm$13 & 0.40$\pm$0.08 \\
\enddata
\tablenotetext{a}{Range of radii of extended X-ray emission.}
\tablenotetext{b}{Orientation of large-scale X-ray structure with
  respect to the nuclear X-ray source.  P.A. and opening angle (half angle).}
\tablenotetext{c}{Net counts and net count rate of extended emission}
\label{tab6.3}
\end{deluxetable}

\clearpage

\begin{planotable}{lcrlcllc}
\small
\tablewidth{6in}
\tablecaption{ {\bf Properties of the Extended X-ray Emitting Gas} }
\tablehead{
\colhead{Galaxy} & \colhead{V} &
\colhead{kT} & \colhead{L$_X$} & \colhead{$f$n$^2$V} &
\colhead{$f^{{{1}\over{2}}}$ n} &
\colhead{$f^{{{1}\over{2}}}$ 2nkT} &
\colhead{$f^{-{{1}\over{2}}}$ t$_{cool}$} \\
\colhead{Name} & \colhead{$\times$ 10$^{65}$} & \colhead{} & 
\colhead{$\times$ 10$^{41}$} &
\colhead{$\times$ 10$^{62}$} &
\colhead{$\times$ 10$^{-2}$} &
\colhead{$\times$ 10$^{-11}$} &
\colhead{$\times$ 10$^{7}$} \\
\colhead{} &
\colhead{(cm$^3$)} & \colhead{(keV)} & \colhead{(erg~s$^{-1}$)} &
\colhead{(cm$^{-3}$)} & \colhead{(cm$^{-3}$)} & 
\colhead{(dyne~cm$^{-2}$)} & \colhead{(yr)} \\
\colhead{(1)} & \colhead{(2)} & \colhead{(3)} & \colhead{(4)} &
\colhead{(5)} & \colhead{(6)} & \colhead{(7)} & \colhead{(8)} \\
} 
\startdata
NGC~2992 & 1.5 & 1.0 & 0.04 & 0.93 & 2.4  &  7.9  & 10 \\
         &     & 0.1 & 0.21 & 5.4  & 5.9  &  1.9  & 0.43 \\
NGC~4388 & 890 & 1.0 & 0.15 & 3.8  & 0.21 &  0.66 & 120 \\
         &     & 0.1 & 0.22 & 5.5  & 0.25 &  0.08 & 10 \\
NGC~5506 & 4.6 & 1.0 & 0.09 & 2.1  & 2.2  &  7.0  & 12 \\
         &     & 0.1 & 0.35 & 8.7  & 4.4  &  1.4  & 0.58 \\
\enddata
\tablenotetext{}{Derived Properties of the X-ray gas for two different
cases: kT $=$ 1.0 keV and kT $=$ 0.1 keV (column 3).
Columns are as follows: (1) name of galaxy, (2) volume of X-ray emitting
gas in units of 10$^{65}$ cm$^3$, assuming the emitting region has a conical
geometry (see text section 4.2),
(3) assumed temperature in keV, (4) 0.2$-$2.4 keV X-ray luminosity 
in units of 10$^{41}$ erg~s$^{-1}$, (5) volume integral of the 
density squared in units of 10$^{62}$ cm$^3$, calculated from the
X-ray luminosity and the ROSAT emissivity from Suchkov et al. 1994,
(6) density $f^{{{1}\over{2}}}$ n in units of 10$^{-2}$ cm$^{-3}$, 
where $f$ is the filling factor of the X-ray gas in the volume, (7) pressure
$f^{{{1}\over{2}}}$ 2nkT in units of 
10$^{-11}$ dyne~cm$^{-2}$, and (8) cooling time of the X-ray gas 
$f$nVkT/L$_X$ in units of 10$^7$ yr.  See text
section 4.2 for additional details.
}
\label{tab7.3}
\end{planotable}

\clearpage

\clearpage

\begin{figure}[p]
\caption{
{\bf NGC~2992. (a, top left) }
ROSAT HRI contours over grayscale plot of B-band GASP image
(from the STScI digitized sky-survey).
The HRI image was smoothed with a Gaussian kernel of $\sigma =$ 24\asec.
{\bf NGC~4388. (b, top right)}
ROSAT HRI contours over grayscale plot of B-band GASP image.
The HRI image was smoothed with a Gaussian kernel of $\sigma =$ 8\asec.
{\bf (c, bottom left)} 
ROSAT PSPC contours over grayscale plot of B-band GASP image.
{\bf NGC~4945. (d, bottom right) }
ROSAT PSPC contours over grayscale plot of B-band GASP image.
} 
\label{xfig1new}
\end{figure}

\begin{figure}[p]
\caption{
{\bf IC~4329A. (a, top left) }
ROSAT PSPC contours over grayscale plot of B-band GASP image.
{\bf NGC~5506. (b, top right) }
ROSAT PSPC contours over grayscale plot of B-band GASP image.
{\bf (c, bottom left)} 
ROSAT HRI contours over grayscale plot of B-band GASP image.
The HRI image was smoothed with a Gaussian kernel of $\sigma =$ 24\asec.
} 
\label{xfig5new}
\end{figure}

\begin{figure}[p]
\caption[Radial Profile of X-ray Emission from NGC~2992]
{ {\bf Radial Profiles of X-ray Emission from NGC~2992.}
Radial profiles in units of counts per 4\asec\ pixel for four quadrants. 
Symbols are as follows:
open triangles, PA $=$ -30\deg\ $-$ 60\deg;
open rectangles, PA $=$ 60\deg\ $-$ 150\deg;
open pentagons, PA $=$ 150\deg\ $-$ 240\deg;
open hexagons, PA $=$ 240\deg\ $-$ 330\deg.
Error bars are 1$\sigma$ of the total number of counts.
{\bf (a)} 20 counts per radial bin.  Note the clear excess emission in
the 2nd quadrant (60\deg\ $-$ 150\deg).
{\bf (b)} 10 counts per radial bin.  Again, the excess emission manifests
itself at the same radii ($\sim$35\asec\ $-$ 45\asec).
} 
\label{xfig9}
\end{figure}
\begin{figure}[p]
\end{figure}
\begin{figure}[p]
\end{figure}
%


\begin{figure}[p]
\caption[Radial Profile of X-ray Emission from NGC~5506]
{ {\bf Radial Profile of Extended X-ray Emission from NGC~5506.}
A model PSF (solid line)
from an observation of HZ43 is also shown for comparison.
} 
\label{xfig11}
\end{figure}
\begin{figure}[p]
\end{figure}
%


\begin{figure}[p]
\caption[Extended radio and X-ray emission in NGC~2992]{ {\bf Extended 
radio and X-ray emission in NGC~2992.} Contours are
1.4 GHz (20~cm) radio continuum emission (see paper~II)
and grayscale
is the ROSAT HRI soft X-ray image (see Figure 1a), 
displayed on
a logarithmic scale to bring out the faint features.  
The lowest surface brightness levels displayed from the X-ray image are
$\sim$1 $\times$ 10$^{-2}$ cnts~s$^{-1}$~arcmin$^{-2}$.
The arrows show the
orientation of the galaxy disk.
Axis coordinates are epoch B1950.
}
\label{ch7fig3}
\end{figure}




\begin{figure}[p]
\caption[Extended radio and X-ray emission in NGC~5506]{ {\bf Extended 
radio and X-ray emission in NGC~5506.} Contours are
soft X-ray emission from the ROSAT HRI image (improved using the
techniques of Morse 1994) and
grayscale is from a 4.9 GHz radio continuum image 
(paper~II).
The greyscale image is displayed on a logarithmic scale to bring out the
structure of the extended radio emission.
The lowest surface brightness levels displayed from the radio image are
$\sim$50 $\mu$Jy~beam$^{-1}$ (beam size 5" FWHM).
The arrows show the orientation of the galaxy disk (horizontal arrows in the 
ps figure are missing).
}
\label{ch7fig4}
\end{figure}
%

\clearpage

%

%
%
%

%

\end{document}